\documentclass[A4]{spie}  
\usepackage[]{graphicx}

\title{Improving the absolute accuracy of the gravitational wave detectors by combining the photon pressure and gravity field calibrators} 

\author{Yuki Inoue\supit{a,b}, Sadakazu Haino\supit{a,b}, Nobuyuki Kanda\supit{c}, Yujiro Ogawa\supit{b,d}, Toshikazu Suzuki\supit{b,e,f}, Takayuki Tomaru\supit{b,d,e,f}, Takahiro Yamamoto\supit{g}, Takaaki Yokozawa\supit{g}
\skiplinehalf
\supit{a}Institute of Physics, Academia Sinica, Taipei 11529, Taiwan; \\
\supit{b}High Energy Accelerator Research Organization (KEK), Ibaraki 305-0801, Japan;\\
\supit{c}Department of Physics, Graduate School of Science, Osaka City University, Osaka 558-8585, Japan;\\
\supit{d}SOKENDAI (The Graduate University for Advanced Studies), Hayama, Miura District, Kanagawa 240-0115, Japan;\\
\supit{e}Institute for Cosmic Ray Research, The University of Tokyo, Chiba 277-8582, Japan;\\
\supit{f}Kavli Institute for the Physics and Mathematics of the Universe (Kavli IPMU), The University of Tokyo, Chiba 277-8568, Japan;\\
\supit{g}Institute for Cosmic Ray Research, The University of Tokyo, Gifu 506-1205, Japan;\\
}

\authorinfo{Yuki Inoue: iyuki@post.kek.jp}

\begin{document} 
\maketitle 

\begin{abstract}
The absolute accuracy of the estimated parameters of gravitational wave sources will be fundamentally limited by the calibration uncertainties of the detectors in upcoming observation runs with the increased number of source statistics.
Photon calibrators have so far been the primary tools for absolute calibration of test-mass displacement, relying on measurement of the photon pressure. 
The current technological limit of the absolute calibration uncertainty for gravitational-wave amplitudes is limited to a few percent, due to the uncertainty in the laser power-standard maintained by the metrology institutes. To reduce this uncertainty, this article proposes a novel calibration method that combines a photon calibrator and a gravity field calibrator. The gravity field calibrator achieves modulation of the displacement of the test mass by generating a gravity gradient. In previous studies, uncertainty in the distance between the test mass and the gravity field calibrator has proven a serious source of systematic error. To suppress this uncertainty, we propose a novel method that uses a combination of quadrupole and hexapole mass distributions in the gravity field calibrator. We estimate the absolute uncertainty associated with method to be as low as 0.17~\%, which is ten times less than that of previous methods.
\end{abstract}



\section{Introduction}

The discovery of gravitational waves (GW) has given us the new probe for observing the universe~\cite{PhysRevLett.116.061102}. 
The typical strain sensitivity, $h$, of second generation interferometric detectors, such as Advanced LIGO~\cite{0264-9381-32-7-074001}, Advanced Virgo~\cite{0264-9381-32-2-024001}, and KAGRA~\cite{0264-9381-29-12-124007, PhysRevD.88.043007}, is around $10^{-23}/\sqrt{\mathrm{Hz}}$ at 100 Hz. 

In GW150914 event data analysis, it has been shown that the calibration errors give significant impact on the sky localization accuracy.
The 90~\% sky confidence region gets larger from $150~\mathrm{deg}^2$ to $610~\mathrm{deg}^2$ by introducing the calibration uncertainties of 10~\% in amplitude and 10 degrees phase~\cite{PhysRevD.93.122004}, and eventually got smaller to $230~\mathrm{deg}^2$ with the improved calibration uncertainties~\cite{PhysRevX.6.041015, PhysRevD.96.102001}.
Using GW signals from compact binary coalescences events, researchers can derive several parameters of the source objects such as masses, spins, luminosity distance, orbital inclination and the sky location. 
The precision of these derived parameters is potentially limited by the calibration accuracy. As the number of detected sources increases and events with higher signal-to-noise ratio (SNR) are detected, calibration uncertainty 
will become the dominant source of errors when extracting physical informations from the signals. 
Testing general relativity has been demonstrated with the GW events from binary black hole mergers~\cite{PhysRevLett.116.221101}. In most of the analysis, the effect of calibration uncertainties on the detection and parameter estimation of GW events have focused on placing constrains on the calibration accuracy by modeling the calibration errors as smooth and random frequency-dependent fluctuations. By a semi-analytic approach to explicitly relate systematic errors in calibration parameters to the GW signal parameters, it has been shown that for events with SNR$\sim20$, calibration accuracy of a few percent is required for certain parameters such as optical gain and actuation strength in order to achieve noise-limited systematics~\cite{Hall:2017off}.
Upper limits and observations of continuous GW waves such as rapidly rotating neutron stars and stochastic background of unresolvable sources depend on calibration uncertainties. The associated uncertainties on the upper limits of continuous waves amount to $\sim20~\%$ by including 10~\% amplitude calibration uncertainty~\cite{0004-637X-839-1-12,PhysRevD.96.122004}.
The suppression of the calibration error also improve burst GW reconstruction.
Especially, the precise correction of the frequency dependence will remove the
biases on arrival time and polarization components. These parameters affects on
estimations for a source direction and rotational axis of Super Novae core, respectively.
In the idea to estimate a mass of an isolated neutron star using gravitational waves~\cite{PhysRevD.91.084032,1742-6596-716-1-012026}, it must determine the phase difference precisely between once and twice spin frequency modes. For many known pulsar cases, these frequencies are around the unity gain frequency where the transfer function phase changes steeply. Also, since the continuous wave measurement would use a long-duration data sets as a order of years, the robustness and stableness of calibrations is essentially important.
In particular, the uncertainty in the absolute amplitude of the GW signal
propagates directly into the estimation of the distance to the sources. 
The rate that compact binary system coalescences in the universe if drawn from detected events. The SNR by the searches are quadratically sensitive to the calibration errors since they are maximized over arrival time, waveform phase, and the template banks. The amplitude calibration uncertainty of 10~\% and the derived uncertainty of the luminosity distances of the sources corresponds to an approximately 30~\% uncertainty in volume and will dominate over the statistical uncertainty~\cite{2041-8205-833-1-L1}.
The detection of a GW signal from the GW170817 Binary Neutron Star (BNS) system, along with a concurrent electromagnetic (EM) signal, began a new era of multi-messenger astronomy~\cite{PhysRevLett.119.161101}. These observations 
allow us to use GW170817 as a standard 
siren~\cite{Abbott:2017xzu,Schutz_1986,Holz_2005,Nissanke_2010} with witch we can 
determine the absolute luminosity distance to the source directly from the 
GW signals. Assuming an event rate of 3000 Gpc$^{-3}$yr$^{-1}$ 
which is consistent with the 90~\% 
confidence interval for GW170817~\cite{PhysRevLett.119.161101}, 
we expect that GW signals will be detected from about 
50 BNS standard sirens dureing the next few observing runs. 
These observations can constrain the Hubble constant ($H_0$) to 2~\% error or less~\cite{Feeney:2018mkj}, and eventually resolve 
the 3-$\sigma$ tension in $H_0$ measurements between Cephied-SN distance 
ladder~\cite{Riess_2016} and CMB data when assuming the $\Lambda$CDM 
model.~\cite{2016-planck} Systematic errors in the calibration of 
the absolute GW signal amplitude must be suppressed less than 1~\% to 
achieve higher-precision $H_0$ measurements using GW standard sirens.



Laser interferometers measure change in distance along the two interferometer arms. Fluctuations in the degree of freedom of the differential arm length (DARM) are suppressed by a DARM control loop. The reconstruction of the DARM fluctuation at the observation frequency is affected by the GWs. The gravitational waveform can be reconstructed from the calibrated error and control signals of this DARM loop. To calibrate these signals, accurate physical models of the actuator and sensing function are essential. These models require measurements of the transfer function and monitoring of the time dependency of the transfer function using continuous sine waves (calibration lines). The residual of the time-dependent model corresponds to the uncertainty of the observaton.

To reduce the systematic uncertainty in the calibration, we need to inject well-parameterized calibration lines for the photon calibrator (Pcal) or other calibration sources for monitoring the time variation of the interferometer responce. The Pcal was developed by the Glasgow and GEO600 reserch groups~\cite{CLUBLEY200185,MOSSAVI20061}, followed by Advanced LIGO which particularly improved Pcals for calibrating the time-dependent response of interferometers~\cite{0264-9381-32-2-024001, doi:10.1063/1.4967303,0264-9381-27-8-084024,0264-9381-26-24-245011,0264-9381-32-2-024001}. However, Pcal still face a challenges in finding of the absolute amplitude calibration because of the uncertainty in the laser power standards published by different national metrology institutes~\cite{EUROMET}. The absolute power between these institutes vary by a few \% \footnote{Figure. 9 at page 46. from \cite{EUROMET} shows the absolute power measurement between the standard institute from nine countries. The systematic discrepancies between nine countries are as large as 3.5 \%.}.

Gravity field calibrator (Gcal) is one of the most promising candidates to be able to solve the uncertainty problem of the absolute laser amplitude calibrations. The technology has been developed and tested by Forward and Miller~\cite{doi:10.1063/1.1709366}, Weber~\cite{PhysRevLett.18.795,PhysRev.167.1145}, University of Tokyo~\cite{Hirakawa,1347-4065-19-3-L123,1347-4065-20-7-L498,PhysRevD.26.729,PhysRevD.32.342} and Rome university group~\cite{Astone1991, Astone1998}. Related techniques using Gcal are discussed in Matone {\it et. al.} and Raffai {\it et. al.}~\cite{0264-9381-24-9-005, PhysRevD.84.082002}. The device can modulate a test mass using a gravity gradient generated by a rotor that depending on the masses, distance, frequency, radius, and the gravity constant. 

This paper proposes new method for achieving sub-percent uncertainty in the absolute amplitude calibration of the GW detectors. The method combines Pcal and Gcal.
Section~\ref{sec:Pcal} explains the methods used for Pcal. In section~\ref{sec:Gcal}, we discuss the principle of a multipole moment of gravity and how it is modulated to derive a calibration signal.
We demonstrate how to calibrate absolute displacement using two calibrators in concert in section \ref{sec:PGCAL}, and in section~\ref{sec:EST}, we discuss the contributions of the systematic error and estimate the current technological limits on the gravitational wave observation from typical physical assumptions. 

\section{Photon calibrator} \label{sec:Pcal}
\begin{figure}
\begin{center}
\includegraphics[width=12cm]{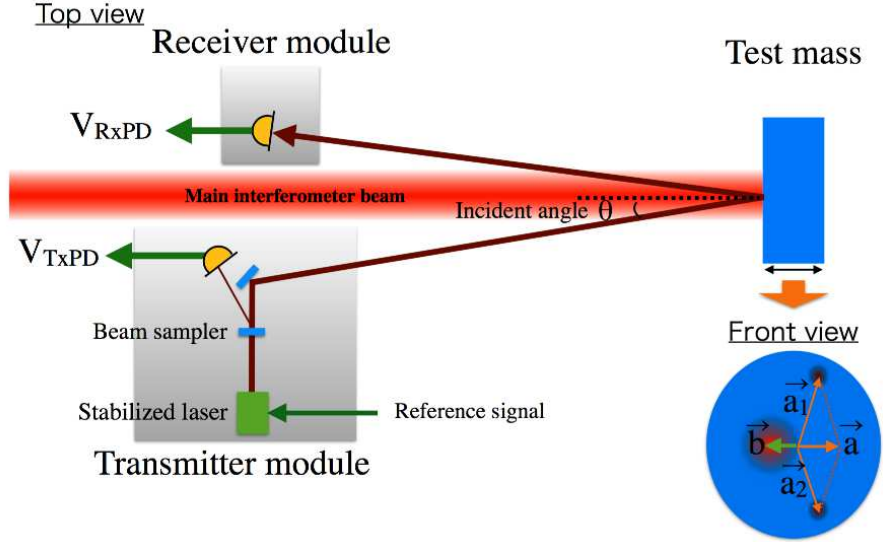}
\caption{Schematic view of photon calibrator. The stabilized laser is placed on the transmitter module. The signal injected to the test mass is monitored through the difference in photo detector responce power between the transmitter module and  receiver module,$V_{TxPD}$  and $V_{RxPD}$. The geometrical factor is characterized in term of the position vectors of the photon calibrator beams, $\vec{a}=\vec{a_1}+\vec{a_2}$, and the main beam, $\vec{b}$.}
\label{fig:Pcal}
\end{center}
\end{figure}

Pcal exploit how the photon radiation pressure from power-modulated laser beams reflects from a test mass. The periodic photon recoil applies a periodic force to test mass~\cite{doi:10.1063/1.4967303}. 
Advanced LIGO, Advanced Virgo and KAGRA employ Pcals for the calibration of the interferometer response~\cite{0264-9381-34-1-015002, KAGRA_Pcal,0264-9381-32-2-024001}. All of them use laser of the same wave length, 1047~nm, to actuate the test mass. The test mass displacement is described as
\begin{equation}
 x = \frac{2P \cos{\theta}}{c} s(\omega)\left(1+\frac{M}{I}\vec{a} \cdot \vec{b} \right) , \label{eq:pcal}
\end{equation}
where $P$ is the absolute laser power, $\theta$ is the incident angle of the Pcal laser, $M$ is the mass of test mass, $\omega$ is the angular frequency of the laser power modulation, and $\vec{a}$ and $\vec{b}$ are the position vectors of the Pcal laser beams. A schematic view of device is shown in Fig.~\ref{fig:Pcal}. $I=Mh^2/12+Mr^2/4$ is the moment of inertia, where $h$ and  $r$ are the thickness and radius of the test mass, respectively. $s(\omega)$ is the transfer function between force and displacement. We can regard the value of $s(\omega)$ as $1/(M \omega^2)$ at the frequency above 20 Hz, as the test mass behaves as a free mass in this regime. 

The amplitude of the laser power noise is stabilized to be less than the design sensitivity. As shown in Fig.~\ref{fig:Pcal}, the power stabilized laser is mounted on the transmitter module. The power of the photo detector responses at the transmitter module, $V_{\mathrm{TxPD}}$, and receiver module, $V_{\mathrm{RxPD}}$ are monitored for differences.  
The largest relative 
uncertainty of photon calibrator is that of laser power.
Advanced LIGO and KAGRA use a working standard to cross-calibrate the relative interferometer responces. The relative uncertainty of each calibrator is 0.51 \%~\cite{doi:10.1063/1.4967303}. 
The second largest relative uncertainty is the optical efficiency of the optical path in the calibrator. We calibrate the injected power from the exterior of the vacuum chamber. Therefore, we need to consider the difference in optical efficiency due to the transmittance of the vacuum window and the reflectance of the mirrors. The measured uncertainty of the optical efficiency in the Advanced LIGO is 0.37 \%. 
For absolute calibration, the photo detector, following the so called gold standard, is calibrated using the laser power standard maintained by the National Institute of Standards and Technology (NIST) in Boulder, CO~\cite{taylor:1994:GEEU} in the U.S. 
The working standard responces for Hanford, Livingston and KAGRA GW detectors are calibrated to this gold standard. 
However, a comparison of the accuracies of the absolute laser power standards maintained by each national standard institute shows a few~\% 
uncertainty~\cite{EUROMET}. 
 This uncertainty leads to the serious systematic error in the distance calibrations propagated from the uncertainty of the absolute calibration.

\begin{table}
\begin{center}
\caption{Specification summary of Advanced LIGO, Advanced Virgo and KAGRA photon calibrator. \label{pcal}}
\footnotesize
\begin{tabular}{cccc}
\hline
& KAGRA& Advanced LIGO& Advanced Virgo \\
\hline
Mirror material & Sapphire & Silica & Silica \\
 Mirror mass & 23 kg & 40 kg & 40 kg \\
  Mirror diameter & 220 mm & 340 mm & 350 mm \\
    Mirror thickness & 150 mm & 200 mm & 200 mm \\
 Distance from Pcal & 36 m & 8 m & 1.5 m \\
to test mass &&& \\
  Pcal laser power & 20 W & 2~W & 3 W \\
  Pcal laser frequency & 1047 nm & 1047 nm &1047 nm\\
  Incident angle& 0.72 deg & 8.75 deg &30 deg \\
  \hline
\end{tabular}
\end{center}
\end{table}

\section{Gravity field calibrator} \label{sec:Gcal}
\begin{figure}
\begin{center}
\includegraphics[width=12cm]{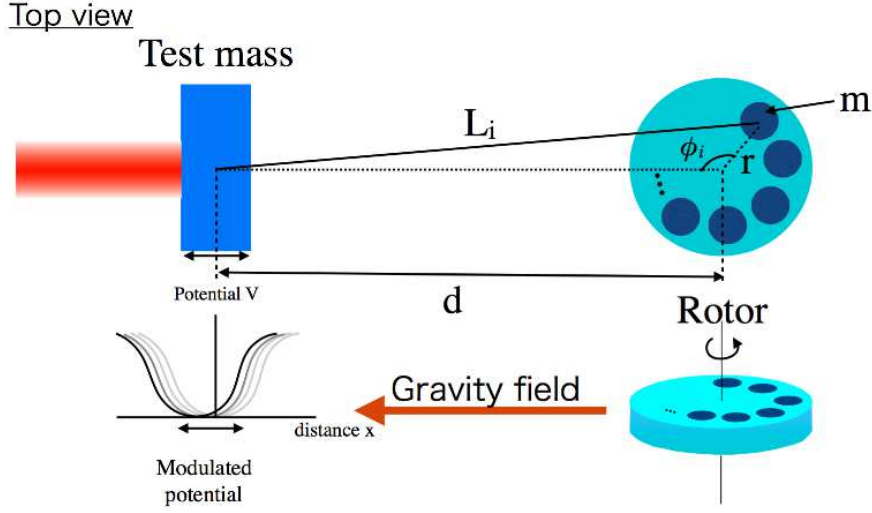}
\caption{Schematic of Gcal. The rotor is placed at  the same height as the test mass and at a distance of $d$. Multipole masses spinning around the rotor generate a varying gravitational potential at the position of test mass.}
\label{fig:dim}
\end{center}
\end{figure}

To address  this problem of uncertainty in the the absolute calibration, we propose a new calibration method that combines Pcal and Gcal. 
 The Gcal generates a dynamic gravity field  by rotating the multipole masses with a rotor placed in a vacuum chamber that isolates acoustic noise. To monitor the frequency of this rotation, an encoder with 
a 16-bit analog to digital converter is included.
Next, we calculate the displacement of the test mass in the dynamic gravity field generated by a multipole moment with N masses.
The calculation assumes a free-mass of a test mass and a set of masses mounted on a disk as shown in Fig ~\ref{fig:dim}.
The rotating the masses $m$ are arranged around the rotor at radius, $r$. 
The distance between the center of this rotor and the test mass mirror is assumed $d$.
We rotate the disk rotates at the angular frequency $\omega_{\mathrm{rot}}=2\pi f_{\mathrm{rot}}$.


We estimate the equation of motion of the test mass as it is moved by the dynamic gravity field.
First, we calculate the distance between the test mass and the N pieces of masses arranged around the rotor.
The distance between i-th mass and the center of test mass is written as
\begin{equation}
L_i=d \sqrt{1+\left( \frac{r}{d} \right)^2 -2\left( \frac{r}{d} \right) \cos{\phi_i} },
\end{equation}
where the angle of the i-th mass is assumed to be $\phi_i=\omega_{\mathrm{rot}} t + 2\pi i/N$.
The gravitational potential at the center of test mass can be described as
\begin{eqnarray}
V &=& \Sigma^N_{i=0} V_i, \\
&=& -GMm \Sigma^N_{i=0}L_i^{-1},\\ \label{eq:vpot}
&=&-\frac{G\!M\!m\!}{d} \Sigma^N_{i=0} \Sigma^{\infty}_{n=0}\! \left(\! \frac{r}{d}\! \right)^n
\!P_n\! \left(\! \cos{\!\left( \! \omega_{\mathrm{rot}} t \!+\!\frac{2 \pi i}{N}\right)\!}\right),
\end{eqnarray}
where $P_n$ is the Legendre polynomial, and $V_i$ is the potential of a mass. The equation of motion of the test mass is 
\begin{eqnarray}
Ma&=&\left| \frac{\partial V}{\partial{d}} \right| =\frac{GMm}{d^2}\Sigma^N_{i=0} \Sigma^{\infty}_{n=0}(n+1) \left( \frac{r}{d} \right)^n \nonumber  P_n\left(\cos{\left(\omega_{\mathrm{rot}} t +\frac{2 \pi i}{N}\right)}\right),
\end{eqnarray}
where $a$ is the acceleration of the test mass. 

\begin{figure}
\begin{center}
\includegraphics[width=12cm]{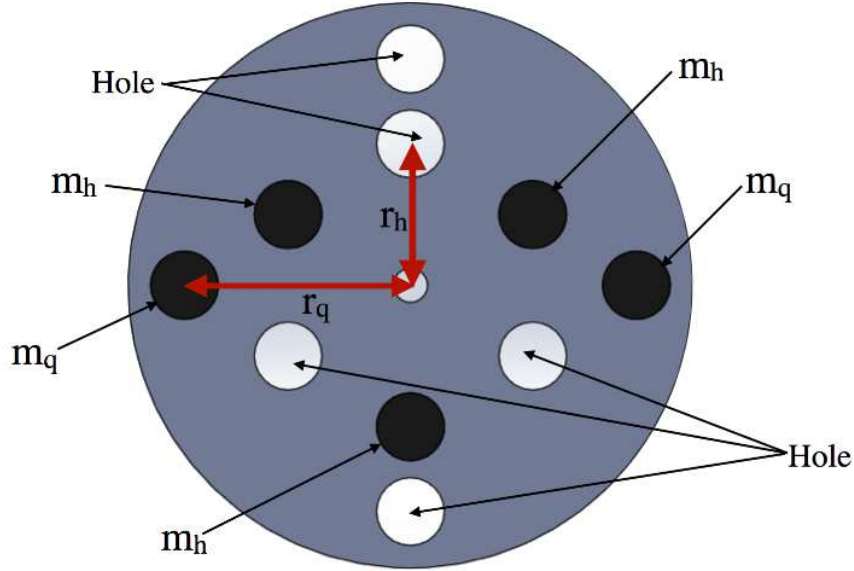}
\caption{Configuration of the rotor with quadrupole and hexapole mass distributions. $m_{\mathrm{q}}$ and $m_{\mathrm{h}}$ are the masses of quadrupole and hexapole masses. $r_{\mathrm{q}}$ and $r_{\mathrm{h}}$ are the radii of the quadrupole and arrangements hexapole.}
\label{fig:hex}
\end{center}
\end{figure}

We arrange the masses around the rotor in a superposition of quadrupole and hexapole arrangements, as shown in Fig.~\ref{fig:hex}. A hole is placed between each mass. These holes is effectively double the magnitude of the gravity gradient. Therefore, the equation of motion of the test mass is 
\begin{eqnarray}
Ma&=&\left| \frac{\partial V}{\partial{d}} \right| =\frac{2GMm}{d^2}\Sigma^N_{i=0} \Sigma^{\infty}_{n=0}(n+1) \left( \frac{r}{d} \right)^n \nonumber P_n\left(\cos{\left(\omega_{\mathrm{rot}} t +\frac{2 \pi i}{N}\right)}\right). \label{eq:EOM}
\end{eqnarray}
Next, we will calculate the displacements of the quadrupole and hexapole rotor masses in sections ~\ref{Quad}  and ~\ref{Hexa}.

\subsection{Displacement of test mass driven by quadrupole mass distribution} \label{Quad}
We calculate the displacement of the quadrupole mass distribution with two pieces and two holes so $N=2$.
The masses and radii of the quadrupole arrangement are $m_{\mathrm{q}}$ and $r_{\mathrm{q}}$. 
The equation of motion for the test mass is
\begin{eqnarray}
Ma&=&\frac{2GMm_{\mathrm{q}}}{d^2}\Sigma^{\infty}_{n=0}(n+1) \left( \frac{r_{\mathrm{q}}}{d} \right)^n \nonumber  \Sigma^1_{i=0}  P_n\left(\cos{\left(\omega_{\mathrm{rot}} t +\pi i \right)}\right).
\end{eqnarray} 
If we assume $r \ll d$, the displacement of the time-dependent lower harmonics can be written as
\begin{equation}
x=\Sigma_{k=1}^{\infty}x_{k\mathrm{f}}\cos(k\omega_{\mathrm{rot}} t)\sim x_{\mathrm{2f}}\cos(2\omega_{\mathrm{rot}} t)=x_{\mathrm{2f}}\cos{\omega t},
\end{equation}
where $k$ is the number of the harmonics. 
The amplitude of the 2-f rotation is then
\begin{equation}
x_{2\mathrm{f}}=9\frac{GMm_{\mathrm{q}}r_{\mathrm{q}}^2}{d^4}s(\omega). \label{2f}
\end{equation}

\subsection{Displacement of test mass driven by hexapole mass distribution} \label{Hexa}
We also calculate the displacement of the hexapole mass distribution with three holes as $N=3$.
The masses and radii of the hexapole distribution are $m_{\mathrm{h}}$ and $r_{\mathrm{h}}$. 
The equation of motion of test mass driven by this arrangement alone is
\begin{eqnarray}
Ma &=& \frac{2GMm_{\mathrm{h}}}{d^2}\Sigma^{\infty}_{n=0}(n+1) \left( \frac{r_{\mathrm{h}}}{d} \right)^n \nonumber  \Sigma^2_{i=0} P_n \left(\cos{\left(\omega_{\mathrm{rot}} t+\frac{2\pi i}{3} \right)} \right).
\end{eqnarray} 
If we assume $r \ll d$, the displacement of the time-dependent lower harmonics can be written as
\begin{equation}
x=\Sigma_{k=1}^{\infty}x_{k\mathrm{f}}\cos(k\omega_{\mathrm{rot}} t)\sim  x_{3\mathrm{f}}\cos(3\omega_{\mathrm{rot}} t)=x_{\mathrm{3f}}\cos{\omega t},
\end{equation}
where amplitude of 3-f is described as
\begin{equation}
 x_{3\mathrm{f}}=15\frac{GMm_{\mathrm{h}}r_{\mathrm{h}}^3}{d^5}s(\omega). \label{3f}
\end{equation}

\section{Absolute power calibration with both photon and Gravity field calibrator} \label{sec:PGCAL}
\begin{figure}
\begin{center}
\includegraphics[width=12cm]{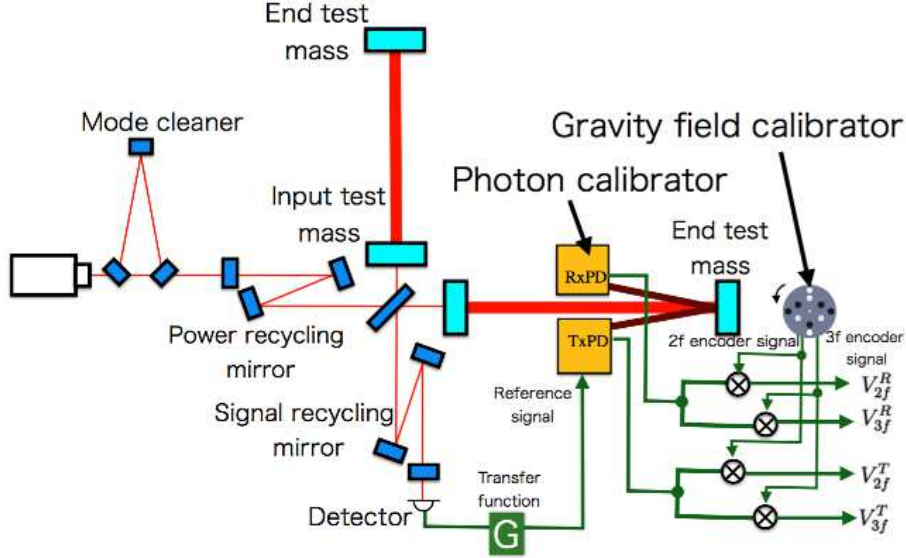}
\caption{Test apparatus for the absolute calibration. The Gcal is placed behind the test mass. The frequency of the Gcal is monitored with the encoder output. The error signal for the differential arm length of the interferometer is sent to the reference port of the photon calibrator for canceling the modulation of the dynamic gravity field with feedback with transfer function $G$. Output signals from the photon calibrator are synchronized with the forces driven by the Gcal. The output signals are demodulated with 2-f and 3-f signals monitored by the encoder.}
\label{fig:IFO}
\end{center}
\end{figure}

This section discusses how to combine the calibration signals from Pcal and Gcal to allow absolute laser power calibration using an interferometer. 
Figure~\ref{fig:IFO} diagrams the combined calibration system.
First, the test mass is driven by the Gcal. The $x_{\mathrm{2f}}$ and $x_{\mathrm{3f}}$ signals are measured from the response of the interferometer. Second, this interferometer signal is sent to the excitation port of the Pcal. This signal acts as a reference signal for feedback control, as shown in Fig.~\ref{fig:IFO}. The Pcal then cancels out the displacement modulated by the Gcal. 
Third, the voltage responses of the transmitter and the receiver module photodetectors are measured. The output signal of the transmitter module, $V_{\mathrm{TxPD}}$ and receiver module, $V_{\mathrm{RxPD}}$ should correspond to the displacement caused by the Gcal. By using Eq~(\ref{eq:pcal}),(\ref{2f}), and (\ref{3f}), the modulated signal powers are
\begin{eqnarray}
 P_{\mathrm{2f}}=\frac{9}{2} \frac{Gcm_{\mathrm{q}}Mr_{\mathrm{q}}^2}{d^4cos\theta}\frac{1}{1+\frac{M}{I}\vec{a}\cdot \vec{b}}, \label{P2f} \\
 P_{\mathrm{3f}}= \frac{15}{2}\frac{Gcm_{\mathrm{h}}Mr_{\mathrm{h}}^3}{d^5cos\theta}\frac{1}{1+\frac{M}{I}\vec{a}\cdot \vec{b}}. \label{P3f}
\end{eqnarray}
Fourth, we demodulate the signal of the transmitter and receiver modules using the measured encoder signal from the Gcal.
The demodulated signals are 
\begin{eqnarray}
V_{\mathrm{2f}}^{\mathrm{T}}=\rho_{\mathrm{T}}P_{\mathrm{2f}}, \\
V_{\mathrm{2f}}^{\mathrm{R}}=\rho_{\mathrm{R}}P_{\mathrm{2f}}, \\
V_{\mathrm{3f}}^{\mathrm{T}}=\rho_{\mathrm{T}}P_{\mathrm{3f}}, \\
V_{\mathrm{3f}}^{\mathrm{R}}=\rho_{\mathrm{R}}P_{\mathrm{3f}}, 
\end{eqnarray} 
where $\rho_{\mathrm{T}}$ and $\rho_{\mathrm{R}}$ are the transfer functions from power to the photo detector output voltages at the transmitter and receiver modules.
Therefore, we can measure the distance from the ratio of responses of the 2-f and 3-f components: 
\begin{equation}
d=\frac{5}{3} \frac{V_{\mathrm{2f}}^{\mathrm{T}}}{V_{\mathrm{3f}}^{\mathrm{T}}}\frac{m_{\mathrm{h}}}{m_{\mathrm{q}}}\frac{r_{\mathrm{h}}^{3}}{r_{\mathrm{q}}^{2}}=\frac{5}{3} \frac{V_{\mathrm{2f}}^{\mathrm{R}}}{V_{\mathrm{3f}}^{\mathrm{R}}} \frac{m_{\mathrm{h}}}{m_{\mathrm{q}}}\frac{r_{\mathrm{h}}^{3}}{r_{\mathrm{q}}^{2}}. \label{d}
\end{equation}

Finally, we calculate the displacement formula for the Pcal calibrated by Gcal. We substitute the Eq. (\ref{2f}) to  Eq. (\ref{eq:pcal}) to obtain the following equation for displacement:
\begin{eqnarray}
x&=&\frac{2P \cos{\theta}}{c} s(\omega)\left(1+\frac{M}{I}\vec{a} \cdot \vec{b} \right), \\
 &=&9\frac{Gm_{\mathrm{q}} M r_{\mathrm{q}}^2}{d^4}\frac{P}{P_{\mathrm{2f}}}s(\omega) , \\
 &=&\frac{729}{625} \frac{GM m^5_{\mathrm{q}} r_{\mathrm{q}}^{10}}{m^4_{\mathrm{h}} r_{\mathrm{h}}^{12} } \frac{{V_{\mathrm{3f}}^{R}}^4}{{V_{\mathrm{2f}}^{R}}^5}V_{\mathrm{in}} s(\omega)  , \label{pcal_new}
\end{eqnarray}
where we assumed that $P(\omega)=\rho_{\mathrm{R}} V_{\mathrm{in}}$, and $V_{\mathrm{in}}$ is the amplitude of the input voltage.
The factor $(GMm_{\mathrm{q}}^5 r_{\mathrm{q}}^{10})/(m_{\mathrm{h}}^4 r_{\mathrm{h}}^{12})$ can be measured before the calibration.  $V^R_{\mathrm{3f}}/V^R_{\mathrm{2f}}$ is measured during the calibration of the Gcal and Pcal. The interval of the calibration signals between the Pcal and Gcal depend on the stability of the photon calibrator laser power. The Advanced LIGO experiment calibrates the absolute laser power using the working standard monthly. Therefore, The Gcal should be run monthly or more frequently. The present method reconstructs the Pcal signal from Gcal signals.  Therefore, the Gcal does not need to be operated during observation runs.
During operation, the Gcal would contaminate the noise floor by adding acoustic and/or vibration noise. However, we can minimize this noise effect by controlling the rotation frequency. The above analysis has not considered to the noise added by the Gcal during observation runs, as we only propose that the Gcal be used to calibrate the absolute displacement before the observations.
The demodulation technique allows us to reduce the systematic error introduced by rotation . When the modulation of Gcal is canceled using  Pcal, the transfer functions of the Gcal and Pcal are also canceled. Therefore, the estimated displacement of the test mass does not depend on the frequency of the rotation. 

\section{Estimation of uncertainty} \label{sec:EST}
In this section, we evaluate the accuracy of the estimated displacement, and discuss the effects on systematic error by changing the operating frequency and distance. After that, we discuss the uncertainty in the displacement of the mirror. 
The following discussion assumes the basic parameters of the KAGRA experiment listed in Table~\ref{pcal}, and the parameters of the Gcal as listed in Table~\ref{sus}. 

\begin{table}
\begin{center}
\caption{\label{sus}Assumed parameters. $G$ is gravity constant~\cite{RevModPhys.88.035009}. $\theta$ is incident angle of the Pcal beams. $M$ is mass of test mass. $1+\frac{I}{M}\vec{a}\cdot \vec{b}$ is geometrical factor.}
\footnotesize
\begin{tabular}{ccc}
\hline
&Value&Relative uncertainty \\
\hline
$G$&$6.67408 \times 10^{-11}~\mathrm{m^3kg^{-1}sec^{-2}}$&0.0047 \%\\
$\cos{\theta}$ &1.000& 0.07~\%\\
$M$ &22.89~kg & 0.02~\%\\
$m_{\mathrm{q}}$&4.485~kg & 0.004~\%\\
$m_{\mathrm{h}}$& 4.485~kg &0.004~\%\\
$r_{\mathrm{q}}$&0.200~m & 0.010~\%\\
$r_{\mathrm{h}}$& 0.125~m & 0.016~\%\\
$1+\frac{I}{M}\vec{a}\cdot \vec{b}$& 1&0.3~\% \\
\hline
\end{tabular}\\
\end{center}
\end{table}

\subsection{Systematic error of higher order terms}
\begin{figure}
\begin{center}
\includegraphics[width=12cm]{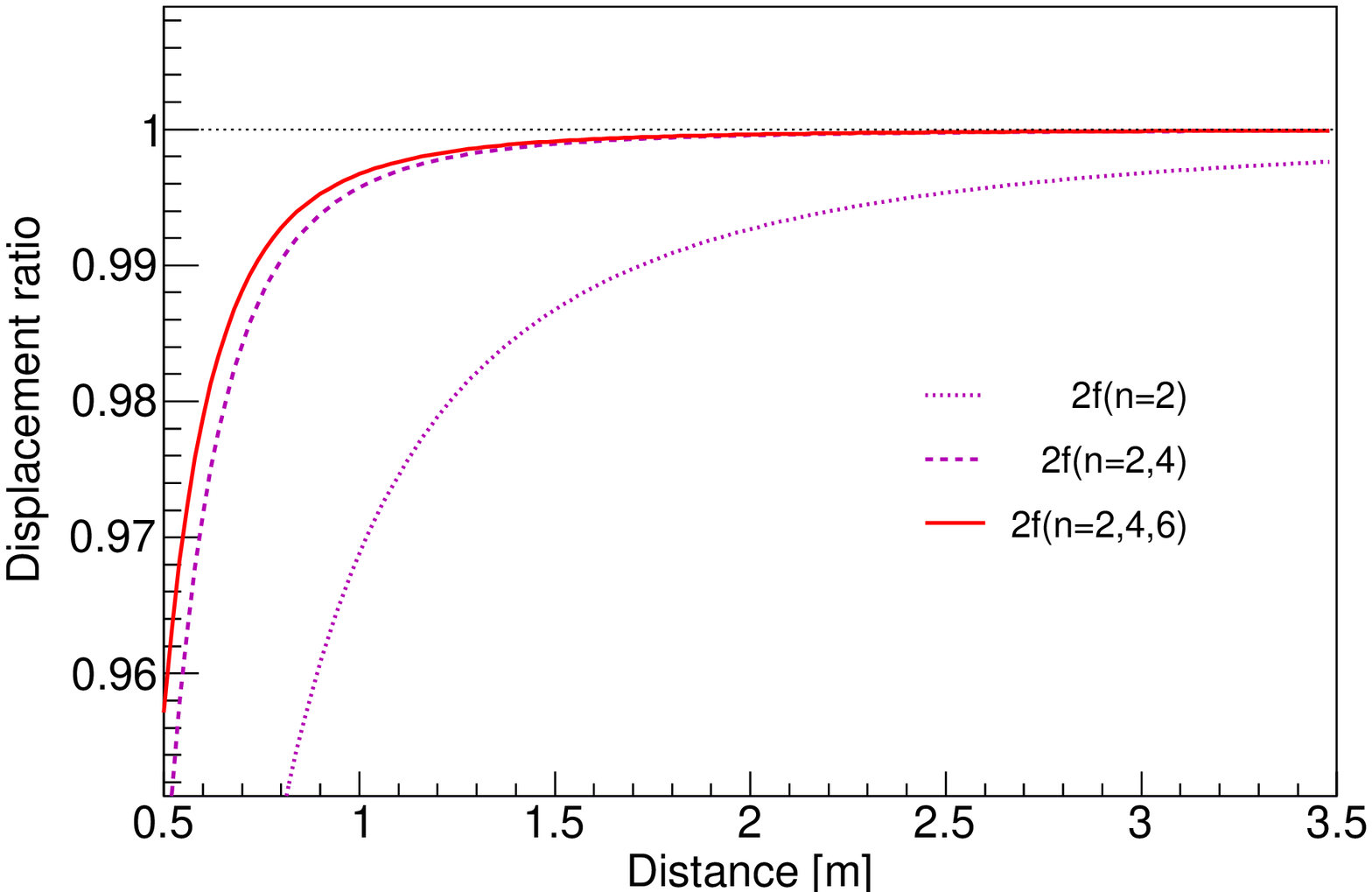}
\caption{Ratio of Legendre polynomial approximations with the numerical calculations of $\frac{\partial V}{\partial{d}}$ and Eq.(~\ref{eq:vpot})
on the test mass displacement for the quadrupole ($N=2$) component as a function of the distance. Dotted, dashed and solid lines correspond to first-order only, second-orders, and third-order approximations, respectively.
The analytical results are listed in Table~\ref{tab:N2}. To achieve precision less than 1~\%, the higher-order terms need to be included.}
\label{fig:FEM-2f}
\end{center}
\end{figure}

\begin{figure}
\begin{center}
\includegraphics[width=12cm]{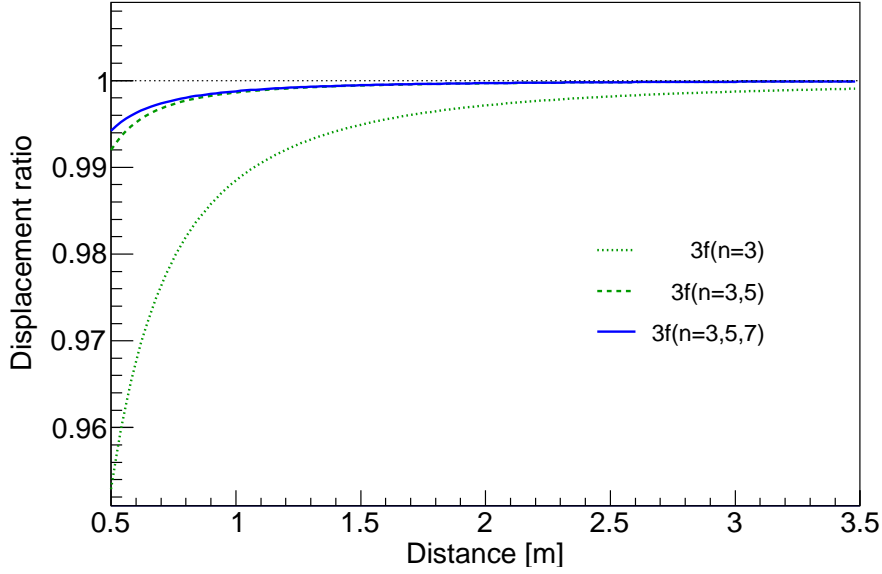}
\caption{Ratio of the Legendre polynomial approximation to the numerical calculations of $\frac{\partial V}{\partial{d}}$ and Eq.(~\ref{eq:vpot}) 
on the test mass displacement for the hexapole distribution ($N=3$) component as a function of distance. The dotted, dashed and solid lines correspond to the first-order only, second-orders, and third-orders, respectively.
The analytical result is listed in Table~\ref{tab:N3}. To achieve the precision less than 1~\%, the higher-order terms need to be included..}
\label{fig:FEM-3f}
\end{center}
\end{figure}

To achieve the precision less than 1 \%, we need to consider the effect of higer-order Legendre polynomials at the position of the test mass. This is because higher-order polynomials also affect the 2-f and 3-f components. The $n$-th order Legendre polynomial is calculated with Eq.(\ref{eq:EOM}). The effect of higher-order factors is mitigated by the factor $(r/d)^n$. Tables~\ref{tab:N2} and \ref{tab:N3} show the calculated displacements of the higher order terms.
To investigate the higher order effects, we compare the estimated test mass displacement between the Legendre polynomial approximation and numerical calculations of $\frac{\partial V}{\partial{d}}$ and Eq.(\ref{eq:vpot}). The ratio of two calculations of the test mass displacement is shown in Figs.~\ref{fig:FEM-2f} and ~\ref{fig:FEM-3f} for the quadrupole ($N=2$) and hexapole ($N=3$) components, respectively, as a function of the distance, $d$.
The results show that the effect of higher-order of polynomials is less than that of systematic error. The mirror therefore needs to be placed at least 2~m away from the rotating mass. Then the sum of the first and second order equations can be used to suppress the systematic error well below 1 \% as shown in Figs.~\ref{fig:FEM-2f} and \ref{fig:FEM-3f}.
If we place the Gcal near the KAGRA end test mass, the distance of 2~m is reasonable. The rotor could be mounted outside of the vacuum chamber. In the following calculations, we assume $d= $~2~m for the simplification of the discussion. 
The analytical calculation of the displacement of the test mass in Eq.(\ref{eq:EOM}) assumes that the rotor masses and the test mass can be approximated as point masses.
We compared the results of this analytical calculation with the numerical integral of the displacements generated by the actual dimensions of the rotor 
with the parameters shown in Table~\ref{sus}, and confirmed that the analytical formula is sufficiently at $d= $~2~m.

\begin{table}
\begin{center}
\caption{Calculated quadrupole ($N=2$) displacement. $n$ is the order of the Legendre polynomial, where $\omega=n\omega_{\mathrm{rot}}$. \label{tab:N2}}
\footnotesize
\begin{tabular}{cccccccc}
\hline
& n=1 & n=2& n=3 &n=4&n=5&n=6&n=7 \\
\hline
1-f&0&0&0&0&0&0&0 \\
2-f&0&$9 \frac{Gmr^2}{d^4\omega^2}$&0&$\frac{25}{4} \frac{Gmr^4}{d^6\omega^2}$&0&$\frac{735}{128} \frac{Gmr^6}{d^8\omega^2}$&0  \\
3-f&0&0&0&0&0&0&0\\
4-f&0&0&0&$\frac{175}{16} \frac{Gmr^4}{d^6\omega^2}$&0& $\frac{273}{32} \frac{Gmr^6}{d^8\omega^2}$&0 \\
5-f&0&0&0&0&0&0&0 \\
6-f&0&0&0&0&0&$\frac{1617}{128} \frac{Gmr^6}{d^8\omega^2}$&0  \\
\hline
\end{tabular}
\end{center}
\end{table}

\begin{table}
\begin{center}
\caption{Calculated hexapole ($N=3$) displacement. $n$ is the order of the Legendre polynomial, where $\omega=n\omega_{\mathrm{rot}}$.  \label{tab:N3}}
\footnotesize
\begin{tabular}{cccccccc}
\hline
& n=1 & n=2& n=3 &n=4&n=5&n=6&n=7 \\
\hline
1-f&0&0&0&0&0&0&0 \\
2-f&0&0&0&0&0&0&0  \\
3-f&0&0&$15\frac{Gmr^3}{d^5\omega^2}$&0&$\frac{315}{32}\frac{Gmr^5}{d^7\omega^2}$&0& $\frac{567}{64} \frac{Gmr^7}{d^9 \omega^2}$\\
4-f&0&0&0&0&0&0&0 \\
5-f&0&0&0&0&0&0&0 \\
6-f&0&0&0&0&0&$\frac{4851}{256} \frac{Gmr^6}{d^8\omega^2}$&0  \\
\hline
\end{tabular}
\end{center}
\end{table}

\subsection{Systematic error of the transfer function}
\begin{figure}
\begin{center}
\includegraphics[width=12cm]{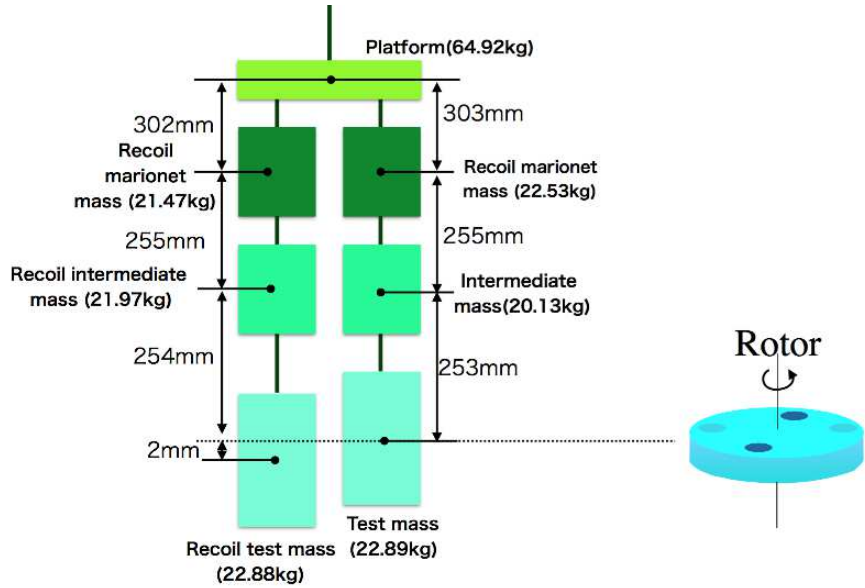}
\caption{Schematic of the suspension system. The parameters of the heights and masses are marked with their assumed values. }
\label{fig:cryo}
\end{center}
\end{figure}

\begin{figure}
\begin{center}
\includegraphics[width=12cm]{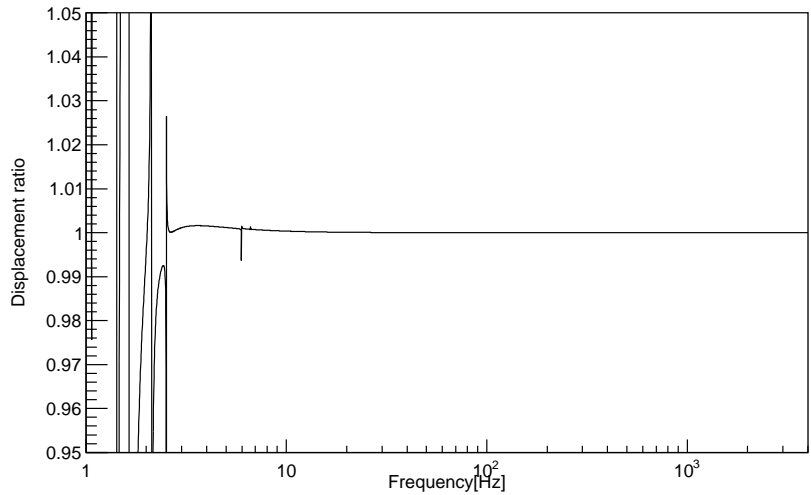}
\caption{Displacement ratio of the transfer function for mult-pendulums by changing modulation frequency. The relations of the modulation frequency, $f$, modulation angular frequency, $\omega$, and rotation angular frequency $\omega_{\mathrm{rot}}$ are described as $n\omega_{\mathrm{rot}}=\omega=2\pi f$. }
\label{fig:ratio}
\end{center}
\end{figure}

The Gcal modulates the test mass mirror with gravitational potential gradient. However, this gradient also actuates the masses of suspension system as shown in Fig.~\ref{fig:cryo}. We simulated the transfer function with the assumption of the cryogenic suspension system installed in KAGRA~\cite{0264-9381-34-22-225001}.The transfer function was calculated using the rigid-body suspension simulation code, called SUMCON~\cite{SUMCON}. We estimated the total displacement by superimposing the displacements driven by both mass distributions. Figure \ref{fig:ratio} shows the displacement ratio between the motion signal and the free-mass motion as a function of frequency. The simulation result is in good agreement with the free-mass motion at this frequencies larger than 20~Hz. The low frequency structures correspond to the resonant peak of the suspension system. Therefore, we can neglect this intermediate-mass effect and regard as motion at frequency over 20~Hz as free-mass motion. 
Therefore, we need to operate the rotor at speeds larger than 20 Hz to achieve error less than 0.1~\%.
We assumed the rotation frequency to be 16~Hz, which corresponds to 32~Hz and 48~Hz for the effective frequency of the 2-f and 3-f components. 
This assumption applies to the discussion in the next section.

\subsection{Uncertainty of displacement and  laser power}
In this section, we estimate the typical displacement based on the result in Table.~\ref{sus}. We neglect Legendre polynomials of degree higher than twwo in the following discussion to simplify the discussion, though they are relevant in the calculations. 
 The estimated 2-f and 3-f displacements are described as
 \footnotesize
\begin{eqnarray}
x^{\mathrm{rms}}_{\mathrm{2f}}&=&1.18 \times 10^{-16}\mathrm{[m]} \times \left( \frac{G}{6.67408 \times 10^{-11} \mathrm{[m^3kg^{-1}sec^{-2}]}} \right) \nonumber \\
&\times&\! \left( \! \frac{m_{\mathrm{q}}}{4.485 \mathrm{[kg]}} \!\right) \! \times \!\left( \!\frac{r_{\mathrm{q}}}{0.200 \mathrm{[m]}} \! \right)^2 \! \times \! \left( \! \frac{2\mathrm{[m]}}{d} \! \right)^4 \! \times \! \left( \! \frac{2\pi \! \times \! 32\mathrm{[Hz]}}{\omega} \! \right)^2,\\
x^{\mathrm{rms}}_{\mathrm{3f}}&=&2.13 \times 10^{-18}\mathrm{[m]} \times \left( \frac{G}{6.6742 \times 10^{-11} \mathrm{[m^3kg^{-1}sec^{-2}]}} \right) \nonumber \\
&\times& \! \left( \! \frac{m_{\mathrm{h}}}{4.485 \mathrm{[kg]}}\! \right) \! \times \! \left(  \!\frac{r_{\mathrm{h}}}{0.125 \mathrm{[m]}} \! \right)^3 
\! \times \! \left(\! \frac{2\mathrm{[m]}}{d} \! \right)^5 \! \times \! \left( \! \frac{2\pi\!\times \! 48\mathrm{[Hz]}}{\omega} \! \right)^2.
\end{eqnarray}
\normalsize

We define the SNR in the term of the ratio of RMS displacement of the design noise spectrum density for the interferometer of KAGRA at 32~Hz for 2-f and 48~Hz for 3-f.
Using this result, we estimate the SNR of the peaks.

\footnotesize
\begin{eqnarray}
\!S\!N\!R_{\mathrm{2f}}&=&392 \times \left(\frac{3.0 \times 10^{-19} [\mathrm{m/\sqrt{Hz}}]}{n_{\mathrm{32Hz}}} \right)  \times \left(\frac{T}{1 [\mathrm{sec}]} \right)^{\frac{1}{2}} \left(\frac{x_{\mathrm{2f}}^{\mathrm{rms}}}{1.178 \times 10^{-16}\mathrm{[m]} }  \right),   \\
\!S\!N\!R_{\mathrm{3f}}&=&73 \times \left(\frac{2.9 \times 10^{-20} [\mathrm{m/\sqrt{Hz}}]}{n_{\mathrm{48Hz}}} \right) \times \left(\frac{T}{1 [\mathrm{sec}]} \right)^{\frac{1}{2}} \left(\frac{x_{\mathrm{2f}}^{\mathrm{rms}}}{2.130 \times 10^{-18}\mathrm{[m] }} \right),
\end{eqnarray}
\normalsize
where $T$ is the integration time. If we integrate a signal larger than 3 min, we can measure $V^{\mathrm{R}}_{2f}$ and $V^{\mathrm{R}}_{3f}$ with sufficiently high SNR that systematic error can be reduced to less than 0.1~\%. 

This method used to measure the absolute laser power as well. The estimated powers are
\footnotesize
\begin{eqnarray}
P_{\mathrm{2f}}&=&0.023 ~\mathrm{[W]}\times \left( \frac{G}{6.6742 \times 10^{-11} \mathrm{[m^3kg^{-1}sec^{-2}]}} \right) \times \left( \frac{m_{\mathrm{q}}}{4.485 \mathrm{[kg]}} \right) \times \left( \frac{r_{\mathrm{q}}}{0.200 \mathrm{[m]}} \right)^2 \times \left( \frac{2\mathrm{[m]}}{d} \right)^4 \nonumber \\ &&\times \left( \frac{1}{\cos{\theta}} \right) \times \left( \frac{1}{1+\frac{M}{I}\vec{a}\cdot \vec{b}} \right)^2,\\
P_{\mathrm{3f}}&=&0.00095~\mathrm{[W]} \times \left( \frac{G}{6.6742 \times 10^{-11} \mathrm{[m^3kg^{-1}sec^{-2}]}} \right) \times \left( \frac{m_{\mathrm{h}}}{4.485 \mathrm{[kg]}} \right) \times \left( \frac{r_{\mathrm{h}}}{0.125 \mathrm{[m]}} \right)^3 \times \left( \frac{2\mathrm{[m]}}{d} \right)^5 \nonumber \\ &&\times \left( \frac{1}{\cos{\theta}} \right) \times \left( \frac{1}{1+\frac{M}{I}\vec{a}\cdot \vec{b}} \right)^2.
\end{eqnarray}
\normalsize
We estimate the laser power with following equations:
\footnotesize
\begin{eqnarray}
\left( \frac{\delta P_{\mathrm{2f}}}{P_{\mathrm{2f}}} \right)^2 &\sim&  \!16 \! \left( \frac{\delta V^{\mathrm{R}}_{{\mathrm{2f}}}}{V^{\mathrm{R}}_{{\mathrm{2f}}}} \!\right)^2+16\left( \frac{\delta V^{\mathrm{R}}_{{\mathrm{3f}}}}{V^{\mathrm{R}}_{{\mathrm{3f}}}} \right)^2 +\left( \frac{\delta P_{\mathrm{sys}}}{P_{\mathrm{sys}}} \right)^2, \label{dP2f} \\ 
\left( \frac{\delta P_{\mathrm{3f}}}{P_{\mathrm{3f}}} \right)^2 &\sim&  \!16 \! \left( \frac{\delta V^{\mathrm{R}}_{{\mathrm{2f}}}}{V^{\mathrm{R}}_{{\mathrm{2f}}}} \right)^2+16\left( \frac{\delta V^{\mathrm{R}}_{{\mathrm{3f}}}}{V^{\mathrm{R}}_{{\mathrm{3f}}}} \right)^2+\left( \frac{\delta P_{\mathrm{sys}}}{P_{\mathrm{sys}}} \right)^2,  \label{dP3f}
\end{eqnarray}
\normalsize
where $\delta P_{\mathrm{sys}}/P_{\mathrm{sys}}$ is the relative systematic error of the power due to the machining tolerance of the rotor masses and radiuses, which are calculated by
\begin{eqnarray}
\frac{\delta P_{\mathrm{sys}}}{P_{\mathrm{sys}}}&\sim& \frac{\delta G}{G} + \frac{\delta M}{M} +\frac{\delta \cos{\theta}}{\cos{\theta}}+ \frac{\delta\left( 1+\frac{M}{I}\vec{a}\cdot \vec{b} \right)}{\! \left( \! 1+\frac{M}{I}\vec{a}\cdot \vec{b} \! \right)}  
+\frac{12}{\sqrt{6}} \frac{\delta r_{\mathrm{h}}}{r_{\mathrm{h}}} +\frac{10}{2} \frac{\delta r_{\mathrm{q}}}{r_{\mathrm{q}}}  +\frac{5}{2} \frac{\delta m_{\mathrm{q}}}{m_{\mathrm{q}}} +\!\frac{4}{\sqrt{6}}  \! \frac{\delta m_{\mathrm{h}}}{m_{\mathrm{h}}}.
\end{eqnarray}
 We next consider the mitigating effect of the systematic error of the masses and radiuses due to the tolerance and uncertainty of the measurement instruments. The values of the masses and radiuses vary slightly with the tolerance of the fabrication process. The errors in $m_{\mathrm{q}}$, $r_{\mathrm{q}}$, $m_{\mathrm{h}}$, and $r_{\mathrm{h}}$ are mitigated by the factor of $1/\sqrt{6}$ and $1/\sqrt{4}$. 
The uncertainty in the quadrupole and hexapole masses are limited by the accuracy of the electronic balance. In this case, we modeled masses made of Tungsten. The density of Tungsten is $19.25~\mathrm{g/cm^3}$. The diameter and thickness of the mass are 0.06~m and 0.08~m, respectively. Therefore, the mass of the rotor mass is 4.485~kg. We assumed that the CG-6000 electronic balance is used to weigh these means, with tolerance of 0.2~g~\cite{CG6000}. Therefore, the relative uncertainty in the mass of the rotor mass is 0.004~\%.

 The rotor disk can be machined by Numerical Control milling. Dimensional accuracy
of less than 0.02 mm can typically be achieved with this process. For measuring the shape, we assume that a three-dimension coordinate measuring machine (CMM) will be employed~\cite{Inoue:16}. The precision of CMM is $2~\mathrm{\mu m}$. This indicates that we can measure the shape of the rotor and masses with  sufficiently low uncertainty using the CMM. 

The estimated relative uncertainties of the laser powers are 0.52~\%. One of the largest uncertainties is the geometrical factor of the Pcal laser. The geometrical factor uncertainty is assumed to be 0.3 \%, which is the same number as the instrument used in Advanced LIGO. 

Finally, assuming that the statistical fluctuations of $V_{in}$, $s(\omega)$, $V^{\mathrm{R}}_{{\mathrm{2f}}}$, and $V^{\mathrm{R}}_{{\mathrm{3f}}}$ are independent for each measurement and therefore can be added in quadrature, the estimated relative uncertainty in the displacement measurements is written as
 \footnotesize
\begin{equation}
\left( \frac{\delta x}{x} \right)^2 \!\sim \! \left( \!\frac{\delta V_{in}}{V_{in}}\! \right)^2+\left(\! \frac{\delta s(\omega)}{s(\omega)} \! \right)^2\!+\!25\!\left(\!\frac{\delta V^{\mathrm{R}}_{{\mathrm{2f}}}}{V^{\mathrm{R}}_{{\mathrm{2f}}}}\! \right)^2+16\!\left(\! \frac{\delta V^{\mathrm{R}}_{{\mathrm{3f}}}}{V^{\mathrm{R}}_{{\mathrm{3f}}}}\! \right)^2\!+ \left(\! \frac{\delta x_{\mathrm{sys}}}{x_{\mathrm{sys}}} \! \right)^2 \label{deltax},
\end{equation}
\normalsize
where $\delta x_{\mathrm{sys}}/x_{\mathrm{sys}}$ is the relative systematic error of the displacement which cannot be added in quadrature. This factor is written as
 \footnotesize
\begin{equation}
\frac{\delta x_{\mathrm{sys}}}{x_{\mathrm{sys}}}=\frac{\delta G}{G} + \frac{\delta M}{M} +\frac{12}{\sqrt{6}} \frac{\delta r_{\mathrm{h}}}{r_{\mathrm{h}}} +\frac{10}{2} \frac{\delta r_{\mathrm{q}}}{r_{\mathrm{q}}}  +\frac{5}{2} \frac{\delta m_{\mathrm{q}}}{m_{\mathrm{q}}} +\!\frac{4}{\sqrt{6}}  \! \frac{\delta m_{\mathrm{h}}}{m_{\mathrm{h}}}.
\end{equation}
\normalsize
We assumed the mitigation factors of radiuses and masses discussed above in this calculation.
To reduce the noise of the displacement measurement, we need to reduce the uncertainty in the shape of the rotor and masses.
The uncertainties in $V^{\mathrm{R}}_{\mathrm{2f}}$,$V^{\mathrm{R}}_{\mathrm{3f}}$, $V^{\mathrm{R}}_{0}$ are much less than that of other contributions. We can reduce the uncertainty of these values using long integration times with statistical measures. Each of the uncertainty is listed in Table~\ref{sus}. The estimated  total uncertainty of the displacement measurement is 0.17~\%.


\section{Conclusion}
Pcal are used in Advanced LIGO, Advanced Virgo and KAGRA. These devices are used to calibrate the interferometer response, and the uncertainty in the calibration affects the estimation of parameters of the GW source. In particular, the distance to the source strongly depends on the absolute laser power of the photon calibrator. In previous studies, the gold standard, in which the interferometer response is calibrated to the NIST laser power standard, was used for the absolute laser power calibration of the photon calibrator. However, the current standard for absolute laser power vary by a few~\% between different countries's metrology institutes. This uncertainty propagates directly to the calculation of the GW detector's absolute displacement.
To address this problem, we proposed a combined calibration method that uses both a Pcal and a Gcal. The Gcal modulated the test mass using a dynamic gravity field. When canceling the displacement of the test mass using the Pcal, the Gcal was used to calibrate the interferometer responce.

This method had the advantage of offering a direct comparison between the amplitudes of the injected power and gravity field modulation at the test mass. Without the proposed gravity-field calibrator, the uncertainties of the optical efficiency through the window and mirrors and the geometrical factor of the laser position need to considered, because the working standard calibration is measured outside of the chamber. However, the method of gravity field can compare the displacement directly. Using this method, the uncertainty of the optical efficiency is avoided when calibrating the absolute laser power. The estimated laser power uncertainty with this method is 0.52~\%. This result suggests that a new power calibration standard can be proposed that gains threefold improvement over the current standards.

Finally, we estimated the uncertainty of absolute calibration with the proposed method. The estimated absolute uncertainty in the displacement measurement is 0.17~\%, which is a tenfold improvement on previous studies. This uncertainty affects the estimation of the distance to the gravitational wave source. This estimated uncertainty brings the precision of the Hubble constant to less than 1~\%. This may address the tension between the Cephied-SN distance ladder~\cite{Riess_2016} and CMB data assuming a $\Lambda$CDM 
model~\cite{2016-planck}.

\acknowledgments     
 
We thank Richard Savage and Darkhan Tuyrnbayev for discussion of the photon calibrator. We would like to express our gratitude to Prof. Takaaki Kajita and Prof. Henry Wong. We would like to thank the KEK Cryogenics Science Center for the support. YI and SH are supported by Academia Sinica and Ministry of Science and Technology (MOST) under Grants No. CDA-106-M06, MOST106-2628-M-007-005 and MOST106-2112-M-001-016 in Taiwan. This work was supported by JSPS KAKENHI Grant Numbers 17H106133 and 17H01135. KAGRA project is supported by MEXT, JSPS Leading-edge Research Infrastructure Program, JSPS Grant-in-Aid for Specially Promoted Research 26000005, MEXT Grant-in-Aid for Scientific Research on Innovative Areas 24103005, JSPS Core-to-Core Program, A. Advanced Research Networks, and the joint research program of the Institute for Cosmic Ray Research, University of Tokyo.

\bibliography{report}   

\begin{thebibliography}{10}

\bibitem{PhysRevLett.116.061102}
Abbott, B.~P. et~al., ``Observation of gravitational waves from a binary black
  hole merger,'' {\em Phys. Rev. Lett.}~{\bf 116},  061102 (Feb 2016).

\bibitem{0264-9381-32-7-074001}
Asai, J. et~al., ``Advanced ligo,'' {\em Classical and Quantum Gravity}~{\bf
  32}(7),  074001 (2015).

\bibitem{0264-9381-32-2-024001}
Acernese, F. et~al., ``Advanced virgo: a second-generation interferometric
  gravitational wave detector,'' {\em Classical and Quantum Gravity}~{\bf
  32}(2),  024001 (2015).

\bibitem{0264-9381-29-12-124007}
Somiya, K., ``Detector configuration of kagra--the japanese cryogenic
  gravitational-wave detector,'' {\em Classical and Quantum Gravity}~{\bf
  29}(12),  124007 (2012).

\bibitem{PhysRevD.88.043007}
Aso, Y., Michimura, Y., Somiya, K., Ando, M., Miyakawa, O., Sekiguchi, T.,
  Tatsumi, D., and Yamamoto, H., ``Interferometer design of the kagra
  gravitational wave detector,'' {\em Phys. Rev. D}~{\bf 88},  043007 (Aug
  2013).

\bibitem{PhysRevD.93.122004}
Abbott, B.~P. et~al., ``Observing gravitational-wave transient gw150914 with
  minimal assumptions,'' {\em Phys. Rev. D}~{\bf 93},  122004 (Jun 2016).

\bibitem{PhysRevX.6.041015}
Abbott, B.~P. et~al., ``Binary black hole mergers in the first advanced ligo
  observing run,'' {\em Phys. Rev. X}~{\bf 6},  041015 (Oct 2016).

\bibitem{PhysRevD.96.102001}
Cahillane, C. et~al., ``Calibration uncertainty for advanced ligo's first and
  second observing runs,'' {\em Phys. Rev. D}~{\bf 96},  102001 (Nov 2017).

\bibitem{PhysRevLett.116.221101}
Abbott, B.~P. et~al., ``Tests of general relativity with gw150914,'' {\em Phys.
  Rev. Lett.}~{\bf 116},  221101 (May 2016).

\bibitem{Hall:2017off}
Hall, E.~D., Cahillane, C., Izumi, K., Smith, R. J.~E., and Adhikari, R.~X.,
  ``{Systematic calibration error requirements for gravitational-wave detectors
  via the Cram{\'e}r-Rao bound},'' (2017).

\bibitem{0004-637X-839-1-12}
Abbott, B.~P. et~al., ``First search for gravitational waves from known pulsars
  with advanced ligo,'' {\em The Astrophysical Journal}~{\bf 839}(1),  12
  (2017).

\bibitem{PhysRevD.96.122004}
Abbott, B.~P. et~al., ``First low-frequency einstein@home all-sky search for
  continuous gravitational waves in advanced ligo data,'' {\em Phys. Rev.
  D}~{\bf 96},  122004 (Dec 2017).

\bibitem{PhysRevD.91.084032}
Ono, K., Eda, K., and Itoh, Y., ``New estimation method for mass of an isolated
  neutron star using gravitational waves,'' {\em Phys. Rev. D}~{\bf 91},
  084032 (Apr 2015).

\bibitem{1742-6596-716-1-012026}
Eda, K., Ono, K., and Itoh, Y., ``Determination of mass of an isolated neutron
  star using continuous gravitational waves with two frequency modes: an effect
  of a misalignment angle,'' {\em Journal of Physics: Conference Series}~{\bf
  716}(1),  012026 (2016).

\bibitem{2041-8205-833-1-L1}
Abbott, B.~P. et~al., ``The rate of binary black hole mergers inferred from
  advanced ligo observations surrounding gw150914,'' {\em The Astrophysical
  Journal Letters}~{\bf 833}(1),  L1 (2016).

\bibitem{PhysRevLett.119.161101}
Abbott, B.~P. et~al., ``Gw170817: Observation of gravitational waves from a
  binary neutron star inspiral,'' {\em Phys. Rev. Lett.}~{\bf 119},  161101
  (Oct 2017).

\bibitem{Abbott:2017xzu}
Abbott, B.~P. et~al., ``{A gravitational-wave standard siren measurement of the
  Hubble constant},'' {\em Nature}~{\bf 551}(7678),  85--88 (2017).

\bibitem{Schutz_1986}
Schutz, B.~F., ``Determining the hubble constant from gravitational wave
  observations,'' {\em Nature}~{\bf 323},  310--311 (Sep 1986).

\bibitem{Holz_2005}
Holz, D.~E. and Hughes, S.~A., ``Using gravitational‐wave standard sirens,''
  {\em The Astrophysical Journal}~{\bf 629},  15--22 (Aug 2005).

\bibitem{Nissanke_2010}
Nissanke, S. et~al., ``Exploring short gamma-ray bursts as gravitational-wave
  standard sirens,'' {\em The Astrophysical Journal}~{\bf 725},  496--514 (Nov
  2010).

\bibitem{Feeney:2018mkj}
Feeney, S.~M. et~al., ``{Prospects for resolving the Hubble constant tension
  with standard sirens},'' (2018).

\bibitem{Riess_2016}
Riess, A.~G. et~al., ``A 2.4
  constant,'' {\em The Astrophysical Journal}~{\bf 826},  56 (Jul 2016).

\bibitem{2016-planck}
Ade, P. A.~R. et~al., ``Planck2015 results,'' {\em Astronomy and
  Astrophysics}~{\bf 594},  A13 (Sep 2016).

\bibitem{CLUBLEY200185}
Clubley, D. et~al., ``Calibration of the glasgow 10 m prototype laser
  interferometric gravitational wave detector using photon pressure,'' {\em
  Physics Letters A}~{\bf 283}(1),  85 -- 88 (2001).

\bibitem{MOSSAVI20061}
Mossavi, K. et~al., ``A photon pressure calibrator for the geo 600
  gravitational wave detector,'' {\em Physics Letters A}~{\bf 353}(1),  1 -- 3
  (2006).

\bibitem{doi:10.1063/1.4967303}
Karki, S. et~al., ``The advanced ligo photon calibrators,'' {\em Review of
  Scientific Instruments}~{\bf 87}(11),  114503 (2016).

\bibitem{0264-9381-27-8-084024}
Goetz, E. et~al., ``Accurate calibration of test mass displacement in the ligo
  interferometers,'' {\em Classical and Quantum Gravity}~{\bf 27}(8),  084024
  (2010).

\bibitem{0264-9381-26-24-245011}
Goetz, E. et~al., ``Precise calibration of ligo test mass actuators using
  photon radiation pressure,'' {\em Classical and Quantum Gravity}~{\bf
  26}(24),  245011 (2009).

\bibitem{EUROMET}
Kuck, S., ``Responsivity of detectors for radiant power of lasers, final
  report, edited by stefan kuck,'' {\em EUROMET Comparison Project}~{\bf 156},
  EUROMET.PR--S2 (2009).

\bibitem{doi:10.1063/1.1709366}
Forward, R.~L. and Miller, L.~R., ``Generation and detection of dynamic
  gravitational‐gradient fields,'' {\em Journal of Applied Physics}~{\bf
  38}(2),  512--518 (1967).

\bibitem{PhysRevLett.18.795}
Sinsky, J. and Weber, J., ``New source for dynamical gravitational fields,''
  {\em Phys. Rev. Lett.}~{\bf 18},  795--797 (May 1967).

\bibitem{PhysRev.167.1145}
Sinsky, J.~A., ``Generation and detection of dynamic newtonian gravitational
  fields at 1660 cps,'' {\em Phys. Rev.}~{\bf 167},  1145--1151 (Mar 1968).

\bibitem{Hirakawa}
H.Hirakawa, K.Tsubono, and K.Oide, ``Dynamical test of the law of
  gravitation,'' {\em Nature}~{\bf 283}(184) (1980).

\bibitem{1347-4065-19-3-L123}
Oide, K., Tsubono, K., and Hirakawa, H., ``The gravitational field of a
  rotating bar,'' {\em Japanese Journal of Applied Physics}~{\bf 19}(3),  L123
  (1980).

\bibitem{1347-4065-20-7-L498}
Suzuki, T. et~al., ``Calibration of gravitational radiation antenna by dynamic
  newton field,'' {\em Japanese Journal of Applied Physics}~{\bf 20}(7),  L498
  (1981).

\bibitem{PhysRevD.26.729}
Ogawa, Y., Tsubono, K., and Hirakawa, H., ``Experimental test of the law of
  gravitation,'' {\em Phys. Rev. D}~{\bf 26},  729--734 (Aug 1982).

\bibitem{PhysRevD.32.342}
Kuroda, K. and Hirakawa, H., ``Experimental test of the law of gravitation,''
  {\em Phys. Rev. D}~{\bf 32},  342--346 (Jul 1985).

\bibitem{Astone1991}
Astone, P. et~al., ``Evaluation and preliminary measurement of the interaction
  of a dynamical gravitational near field with a cryogenic gravitational wave
  antenna,'' {\em Zeitschrift f{\"u}r Physik C Particles and Fields}~{\bf 50},
  21--29 (Mar 1991).

\bibitem{Astone1998}
Astone, P. et~al., ``Experimental study of the dynamic newtonian field with a
  cryogenic gravitational wave antenna,'' {\em The European Physical Journal C
  - Particles and Fields}~{\bf 5},  651--664 (Oct 1998).

\bibitem{0264-9381-24-9-005}
Matone, L. et~al., ``Benefits of artificially generated gravity gradients for
  interferometric gravitational-wave detectors,'' {\em Classical and Quantum
  Gravity}~{\bf 24}(9),  2217 (2007).

\bibitem{PhysRevD.84.082002}
Raffai, P., Szeifert, G., Matone, L., Aso, Y., Bartos, I., M\'arka, Z., Ricci,
  F., and M\'arka, S., ``Opportunity to test non-newtonian gravity using
  interferometric sensors with dynamic gravity field generators,'' {\em Phys.
  Rev. D}~{\bf 84},  082002 (Oct 2011).

\bibitem{0264-9381-34-1-015002}
Tuyenbayev, D. et~al., ``Improving ligo calibration accuracy by tracking and
  compensating for slow temporal variations,'' {\em Classical and Quantum
  Gravity}~{\bf 34}(1),  015002 (2017).

\bibitem{KAGRA_Pcal}
Inoue, Y. et~al. {\em In preparation}  (2018).

\bibitem{taylor:1994:GEEU}
Taylor, B.~N. and Kuyatt, C.~E., ``Guidelines for evaluating and expressing the
  uncertainty of nist measurement results,'' tech. rep., NIST Tecnical Note
  1297 (1994).

\bibitem{RevModPhys.88.035009}
Mohr, P.~J., Newell, D.~B., and Taylor, B.~N., ``Codata recommended values of
  the fundamental physical constants: 2014,'' {\em Rev. Mod. Phys.}~{\bf 88},
  035009 (Sep 2016).

\bibitem{0264-9381-34-22-225001}
Michimura, Y. et~al., ``Mirror actuation design for the interferometer control
  of the kagra gravitational wave telescope,'' {\em Classical and Quantum
  Gravity}~{\bf 34}(22),  225001 (2017).

\bibitem{SUMCON}
SUMCON {\em
  https://gwdoc.icrr.u-tokyo.ac.jp/cgi-bin/DocDB/ShowDocument?docid=3729} .

\bibitem{CG6000}
CG-6000 {\em http://www.vibra.co.jp} .

\bibitem{Inoue:16}
Inoue, Y. et~al., ``Two-layer anti-reflection coating with mullite and
  polyimide foam for large-diameter cryogenic infrared filters,'' {\em Appl.
  Opt.}~{\bf 55},  D22--D28 (Dec 2016).

\end{thebibliography}
\bibliographystyle{spiebib}   

\end{document}